\documentclass[preprint,aps,prc,showpacs,nofootinbib]{revtex4-1}
\usepackage{amsmath}
\usepackage{amssymb}
\usepackage{graphics}

\usepackage{epsfig}

\newcommand{\Li}{\mbox{Li}_2}

\newcommand\ba{\begin{eqnarray}}
\newcommand\ea{\end{eqnarray}}
\newcommand\be{\begin{equation}}
\newcommand\ee{\end{equation}}
\newcommand\nn{\nonumber}
\newcommand{\bas}{\begin{eqnarray*}}
\newcommand{\eas}{\end{eqnarray*}}

\begin{document}

\title{Compilation and analysis of charge asymmetry measurements from electron and positron scattering on nucleon and nuclei}

\author{E.~Tomasi-Gustafsson}
\email{Corresponding author: etomasi@cea.fr}
\affiliation{\it CEA,IRFU,SPhN, Saclay, 91191 Gif-sur-Yvette Cedex, France, and \\
CNRS/IN2P3, Institut de Physique Nucl\'eaire, UMR 8608, 91405 Orsay, France}
\author{M. Osipenko}
\affiliation{\it I.N.F.N., via Dodecaneso 33, Genova, I-16146, Italy}
\author{E.~A.~Kuraev, Yu.~Bystritsky}
\affiliation{\it JINR-BLTP, 141980 Dubna, Moscow region, Russian Federation}


\begin{abstract}
World data on the lepton-charge asymmetry in the elastic and inelastic lepton scattering off the proton and nuclei are compiled and discussed. After reviewing the published results, we compare the elastic data to a model calculation of the two-photon exchange mechanism. We show that the existing data do not provide any evidence for the two-photon contribution. At significance level 0.05 the data allow to exclude the two-photon exchange as an explanation for the difference between Rosenbluth and polarization measurements of proton electromagnetic form factors.
\end{abstract}

\maketitle
\section{Introduction}

Although the elastic electron-hadron scattering is one of the simplest elementary reactions, it is an object of large experimental and theoretical effort since many decades. Most of the interpretation of the observables, in polarized and unpolarized scattering, is based on the assumption that the interaction of the lepton with a hadron (proton or nucleus):
\be
e(k)+h(p)\to e(k')+h(p'),
\label{eq:eqreac}
\ee
(in brackets are the four momenta of the corresponding particles) occurs through the exchange of one virtual photon, with four-momentum $q=k-k'$ ($q^2<0$, and $-q^2=Q^2$). This is so called One-Photon-Exchange (OPE) approximation or Born approximation.

Since early sixties it was noted in the literature that the Two ($n$)-Photon Exchange (TPE) could also contribute to the observables, although the size of its amplitude would be scaled by the factor $Z\alpha$ $((Z\alpha)^n)$ (where $\alpha=1/137$ is the fine structure constant of the electromagnetic interaction, and $Z$ is the target charge number).
In order to find experimental evidence for the processes beyond OPE one has to base on various theoretical predictions.
First of all TPE is proportional to the target charge $Z$, hence a nuclear target is desirable.
In Ref.~\cite{twof} it was suggested that a possible large effect from the TPE could arise at large $Q^2$. In this kinematics the factor $\alpha$ may be compensated by the steep increase of proton form factors (FFs), if the transferred momentum is equally shared between the two photons. This effects becomes important for large target mass $A$ as the $Q^2$-slope of FFs depends on the number of constituents. From quark counting rules, asymptotically, one expects FFs to decrease as ${(Q^2)}^{-(n-1)}$, with $n=3(6)$ for the proton (deuteron)~\cite{QCD}.

Finally, in frame of the model~\cite{Bl05} and in QED calculations~\cite{Ku06} it is found that the TPE contribution is larger at backward angles.

Suggested processes for the search of TPE contribution are the elastic lepton scattering off the proton or nuclei and the crossed channels (annihilation processes $e^+ + e^-\to p+\bar p$ and $p+\bar p\to e^+ + e^-$).
The unpolarized elastic scattering provides signatures which give  model independent information on the presence of TPE:
\begin{itemize}
\item non-linearities in the Rosenbluth plot, i.e., in the reduced cross section versus $\epsilon$ at fixed $Q^2$, where
$\epsilon^{-1}=1+2(1+\tau)\tan^2(\theta/2)$ is the linear polarization of the virtual photon, $\tau={Q^2}/{(4M^2)}$, $M$ is the proton mass, and $\theta$ is the angle of the scattered electron in the laboratory (lab) reference frame.
\item non-vanishing lepton-charge asymmetry which is defined as:
\be
A^{odd}=
\displaystyle\frac{\sigma (e^- p\to e^-p )-\sigma (e^+ p\to e^+p )}
{\sigma (e^- p\to e^-p )+\sigma (e^+ p\to e^+p )}.
\label{eq:eqasym}
\ee
\end{itemize}
In the crossed channels, TPE can be searched in the angular distribution, in particular through an odd contribution with respect to $\cos\tilde{\theta}$, where $\tilde{\theta}$ is the center of mass (cms) scattering angle of the produced particle (see Refs.~\cite{Re99,ETG08} and References therein).

The history of experimental studies related to TPE mechanism can be briefly summarised as follows:
\begin{itemize}
\item In the 70's the presence of a possible TPE contribution was object of  extended experimental and theoretical investigations in unpolarized lepton scattering off the proton and nuclei. As a conclusion of a series of measurements, detailed below, no experimental evidence was found and since that time the OPE approximation was assumed {\it a priori}.

\item A dedicated $ed$ scattering experiment was performed in Ref.~\cite{Pr68} to search for the recoil deuteron vector polarization. A non-zero polarization would provide an evidence for TPE contribution. However, the measured value was compatible with zero.

\item In 1998~\cite{Re99} TPE was suggested as a possibility to reconcile two sets of data on electron deuteron elastic scattering. Since the experimental techniques allowed to apply the polarization method \cite{Re67}, new precise data on the ratio of the electric to magnetic FFs were obtained and a discrepancy in the ratio of the electric and magnetic FFs of the proton, measured by polarized and unpolarized experiments was observed (for a review, see Ref. \cite{CFP08}). 

\item The analyses of the elastic cross section data in terms of deviation from the linearity in the Rosenbluth plot~\cite{ETG05,Tv06}, did not show any evidence for the TPE contribution. In Ref.~\cite{ETG05} it was also pointed out that radiative corrections (RC), as they were applied to the data, may induce important effects on the relevant observables. In particular RC change dramatically the slope of the Rosenbluth plot, and even sign of the slope, for  $Q^2> 2$ GeV$^2$.

\item Recently, the GEp collaboration measured the angular dependence of the ratio of longitudinal to transverse polarization, more exactly its $\epsilon$-dependence \cite{Meziane:2010xc}. The very precise results, showing a constant behaviour as a function of $\epsilon$, are in agreement with the OPE expectation and with Ref. \cite{By07}.

\item Due to the lack of statistics, few data exist on angular distributions in the annihilation region. Recently the process $e^++e ^-\to  p+\bar p +\gamma$ has been measured by the BABAR collaboration \cite{Au05}. The initial state radiation, when the photon is sufficiently hard, allows to factorise out the kinematic terms associated to the photon and to extract the differential cross section of the process $e^++e ^-\to p+\bar p$. The analysis of these data in terms of angular asymmetry was done in Ref.~\cite{ETG08} and showed no visible TPE effect in the limit of the errors.

\item Results from the HERMES collaboration on deep inelastic scattering showed no observable single spin asymmetries in electron and positron scattering on a transversely polarized proton target
\cite{Airapetian:2009ab}.

\item A re-analysis of a selected sample of the existing data concluded in evidence for TPE contribution~\cite{Ar04}, and that TPE suppresses the cross section at low $\epsilon$ and low $Q^2$. In Refs.~\cite{Al09,Ch08} predictions were done for charge asymmetry measurements, under specific assumptions and parametrization of the TPE contribution.

\end{itemize}

The purpose of this work is to collect and discuss the world data on lepton-charge asymmetry from elastic and inelastic scattering on proton and nuclei, and their dependencies on the relevant kinematic variables. The elastic scattering data are compared point by point with a recent first order calculation, Ref.~\cite{Ku08}. A good agreement appears between this calculation and the data, in the limit of the experimental errors.

\section{Definitions}

In this work we focus our attention on an observable which is sensitive to the real part of the TPE amplitude: the ratio of the elastic lepton-nucleon scattering for positive and negative charged leptons.
In Born approximation, which corresponds to the lowest order diagram for OPE, the elastic lepton-nucleon scattering is symmetric with respect to the lepton charge sign. As indicated above, the presence of TPE, more exactly the interference between OPE and TPE (that we will note as $TPE$ for simplicity), induces lepton-charge odd (C-odd) contributions in the matrix element. 

Let us stress that C-odd terms arise also from other radiative correction contributions and from $Z$-boson exchange (which is negligible in the kinematical range considered in the present work). C-odd terms at order $\alpha^3$ are induced both from TPE and from the interference of real photon emission from lepton and proton. Indeed, for high energy experiments where $Q^2\gg M^2$ the proton bremsstrahlung should be taken into account. 

The evaluations of proton bremsstrahlung and TPE are model dependent as they contain information on the proton structure. In the corrections applied to the experimental data, based mostly on Refs \cite{Me63,Mo69,MT00} the finite part of the TPE is usually neglected. These two types of contributions contain infrared divergences which cancel in their sum and can not be considered separately. In order to extract the part of the C-odd corrections included in the experimental results, below we define the different terms of radiative corrections as follows:
\begin{itemize}
\item {\it soft photon contribution} refers to those terms related to the real soft photon bremsstrahlung, after the cancellation of infrared divergences. This contribution depends on the maximum energy, $\Delta E$, of the photon that escapes the detection;
\item {\it hard box terms} are due to TPE corrections (excluding the divergent part).
\end{itemize}
The results from Ref.~\cite{Ku08} show that the lepton-charge asymmetry may be large due to the soft photon terms, for small values of $\Delta E$, while the hard box terms are small in all the investigated kinematic domain.

A model independent calculation of the lepton-charge asymmetry is possible only if the target is structure less, as $\mu$ or $e$~\cite{Ku06}. In Ref.~\cite{Kuraev:2008gw} it is suggested that the TPE contribution calculated for $e+\mu$ scattering corresponds to an upper limit (in absolute value) for complex target. The reason is that proton form factors are smaller than unity in the relevant $Q^2$ range. Inclusion of nucleon excitations shows a compensation between inelastic and elastic intermediate states. Such indications are based on model calculations~\cite{By07,Ko07}, as well as on analyticity arguments~\cite{Ku08}. Bounds to charge asymmetry were already derived in Ref. \cite{DeRujula:1973pr}.

The unpolarized cross section $d\sigma_B$, for lepton-hadron elastic scattering at the lowest order of perturbation theory, assuming OPE, can be expressed in terms of two structure functions, $A$ and $B$, which depend only on the square momentum of the transferred photon, $Q^2$:
\be
d\sigma_B (e^{\pm} h\to e^{\pm}h )=
d\sigma _{Mott}
\left [A(Q^2)+B(Q^2) \tan ^2\frac{\theta}{2}\right ],
\label{eq:eqs}
\ee
where $d\sigma_{Mott}$ is the cross section for point-like particles. This is a very general expressions that holds for any hadron of any spin $S$. The structure functions depend on the $2S+1$ electromagnetic form factors of the hadron. In the Born approximation, the elastic cross section is identical for positrons and electrons. A deviation of the ratio
\be
R=
\displaystyle\frac{\sigma (e^+ h\to e^+h )}
{\sigma (e^- h\to e^-h )}=\frac{1-A^{odd}}{1+A^{odd}}
\label{eq:ratio}
\ee
from unity would be a clear signature of C-odd contributions to the cross section.
In Ref.~\cite{Ku08}, an exact QED calculation was performed for $e^{\pm}\mu$ scattering, and the crossed process.
The obtained lepton-charge asymmetry at first order in $\alpha$\footnote{Note a difference of sign in the definition, Eq.~(2) in Ref.~\protect\cite{Ku08}.} is given by:
\ba
A^{odd}&=&\frac{d\sigma^{e^-p}-d\sigma^{e^+p}}{2d\sigma^B (1+\delta^{even})}
 =\frac{2\alpha}{\pi (1+\delta^{even})}
\biggl[ \ln\rho\ln\frac{(2\Delta E^\prime)^2}{ME}
+\frac{5}{2}\ln^2\rho
-\ln x\ln\rho \nn \\
&&
-\Li\left (1-\frac{1}{\rho x}\right)
+\Li\left(1-\frac{\rho}{x}\right )\biggr ],
~\Li\left (z \right)=-\int_0^z\frac{dx}{x}\ln(1-x),
\label{eq:eqv1}
\ea
with
$$
\rho =\left (1-\frac {Q^2}{s}\right)^{-1}=1+2\frac{E}{M}\sin^2\frac{\theta}{2},
~x=\frac{\sqrt{1+\tau}+\sqrt{\tau}}{\sqrt{1+\tau}-\sqrt{\tau}},
$$
$E$ is the initial energy, $\rho$ is the inverse fraction of the initial energy carried by the scattered electron, $\rho=E/E'$, $\delta^{even}$ is the C-even RC factor~\cite{MT00}
and $\Delta E^\prime$ is the maximum energy of the photon that can escape the detection. It has to be noted here that the definition of $\Delta E^\prime$ may lead to ambiguities. For example, assume one measures the scattered lepton alone and apply a cut on the scattered lepton energy $E^\prime > E_3-\Delta E^\prime$ to select the elastic peak. Then one can miss a photon with energy $\Delta E^\prime$ emitted from the scattered electron, but for the photon emitted from the initial lepton the allowed energy will be $\Delta E=\rho^2 \Delta E^\prime$.
More complex variations are possible in coincidence experiments, detecting both scattered lepton and recoiled proton.

The asymmetry above was expressed as the sum of the contribution from TPE, (more exactly the interference between the Born amplitude and the box-type amplitude) and terms from soft photon emission. The latter term gives the largest contribution to the asymmetry and contains a large $\epsilon$ dependence.

\section{Compilation of $e^{\pm}+ p$ scattering data}

\subsection{Elastic scattering}

\begin{figure}
\begin{center}
\includegraphics[bb=1cm 6cm 20cm 23cm, scale=0.4]{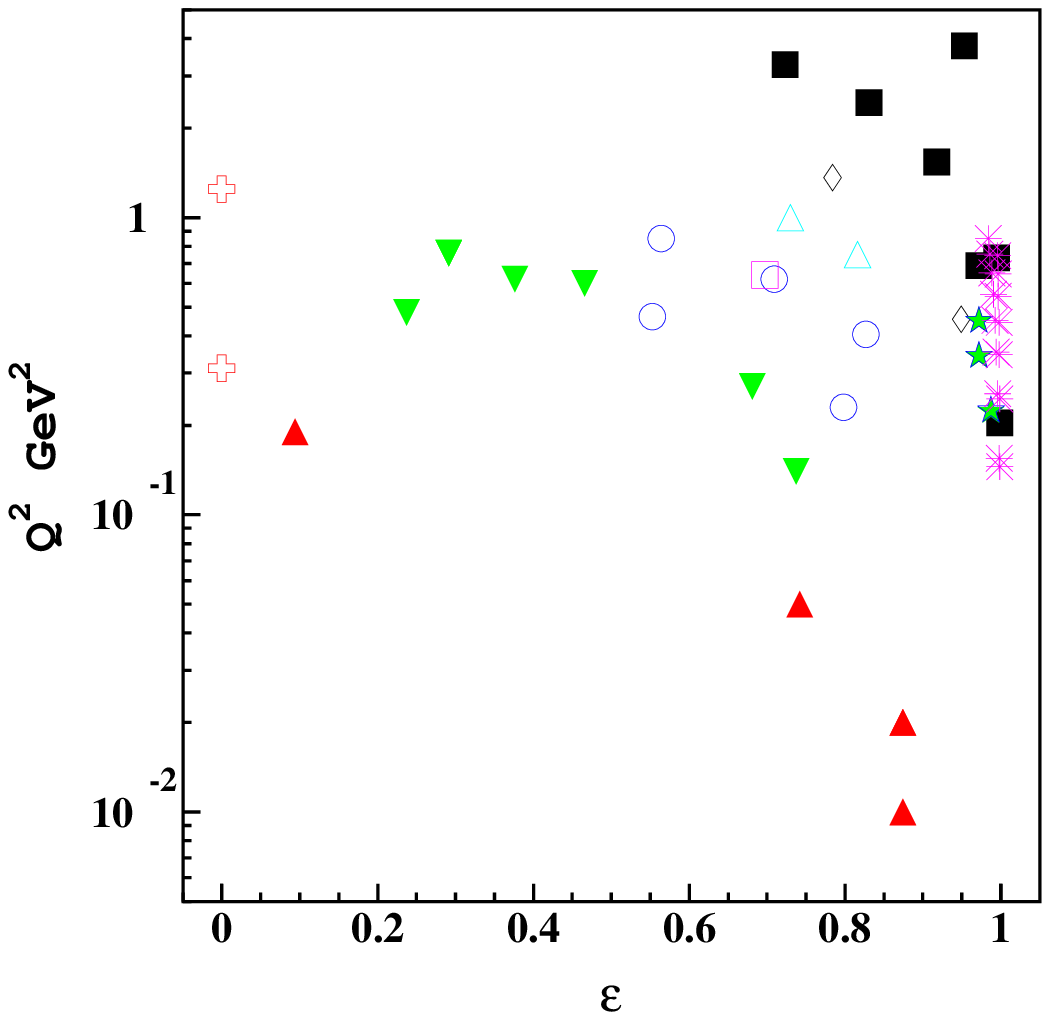}~
\includegraphics[bb=1cm 6cm 20cm 23cm, scale=0.4]{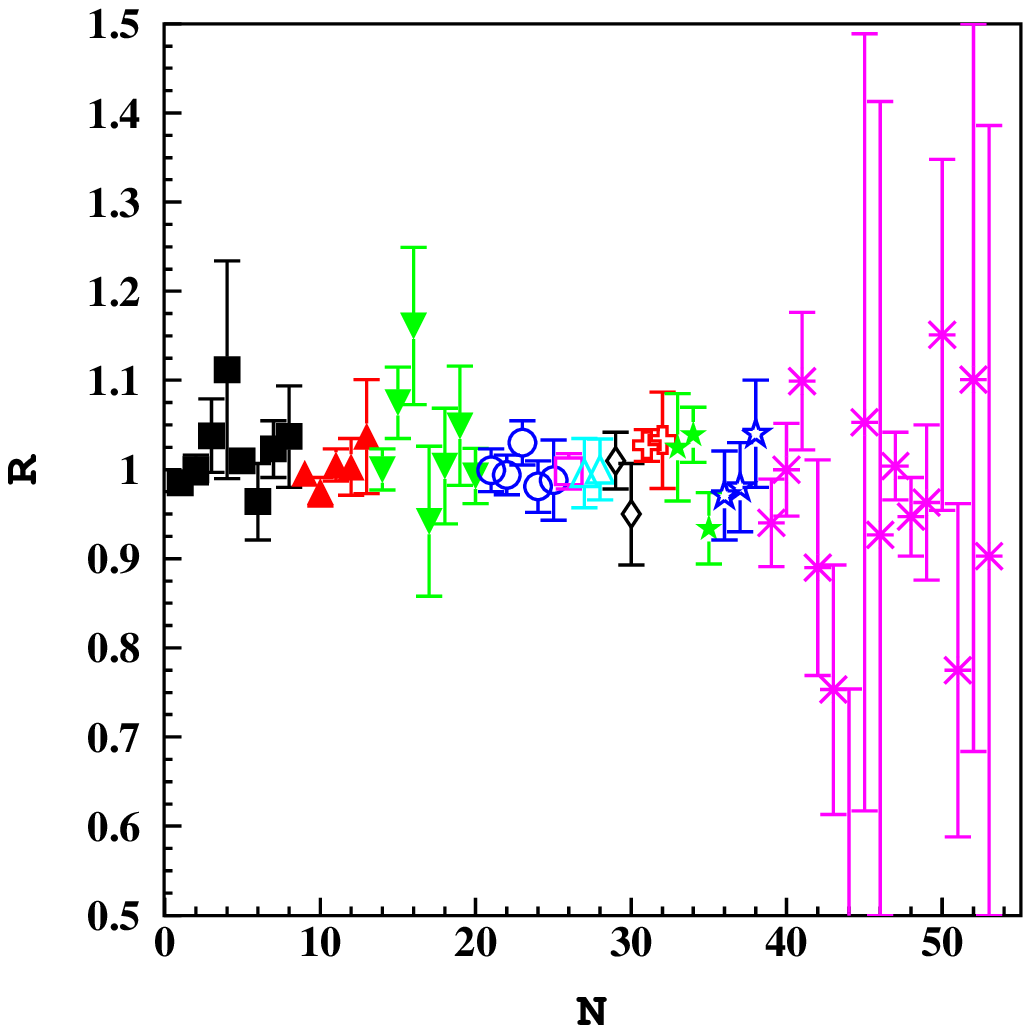}
\caption{\label{fig:r_el_data} (Color online) Kinematics and absolute values of the published elastic cross section ratios,  $R=\sigma(e^+p)/\sigma(e^-p)$. Here $N$ is a number attributed to each point, following a sequential order from Refs.
\protect\cite{Mar68} (solid squares, black),
\protect\cite{Yo62} (triangles, magenta),
\protect\cite{Br65} (triangles down, green),
\protect\cite{An68} (open circles, blue),
\protect\cite{An66} (open square, magenta),
\protect\cite{Ba67} (open triangles, cyan),
\protect\cite{Ca67} (open lozenges, dark blue)
\protect\cite{Bo68} (open crosses, red),
\protect\cite{Ha79} (stars),
\protect\cite{Ha75} (open stars, blue),
\protect\cite{Ca69} (asterisks, magenta).}
\end{center}
\end{figure}

The unpolarized cross section of electron and positron scattering on hadronic targets was extensively studied in the 70's, in dedicated experiments.
The world data on the lepton-charge asymmetry in the elastic scattering off the proton target as well as on nuclei, are summarised in Fig.~\ref{fig:r_el_data}. Most of the data concern electron and positron beams, few data correspond to muon beams. Different measurement techniques were used. The simpler measurements were performed in Refs.~\cite{Yo62,Mar68,Ca69,Ha75,Ha76,Ha79,Ba67} and experiment II of Ref.~\cite{Br65} where a single arm magnetic spectrometers detects the scattered lepton. The second type of measurements was based on the detection of the scattered lepton and the recoil proton in coincidence as in Refs.~\cite{An66,Ca67,An68} and experiment I of Ref.~\cite{Br65}. In these experiments, only the angles of the two particles were measured. The cut on the difference between initial beam energy and that reconstructed from the two angles allowed to select elastic scattering channel. A particular technique was used in Ref.~\cite{Bo68}, where both scattered lepton and recoiled proton were measured with magnetic spectrometers. Various radiative corrections had been applied to the data. In order to have a coherent picture of entire database we removed the original radiative corrections from the published data and calculated our theoretical expectation $R^{th}$ from Eq.~\ref{eq:eqv1}. This task was sometimes nontrivial due to lack of information or different event topology as discussed below. In the cases of coincidence measurements, in principle, a complete calculation of radiative corrections should be done on the basis of a five-fold radiative cross-section and embedded in a dedicated simulation programme for each experiment, what is out of reach for most of the past experiments. Therefore, we apply Eq.~\ref{eq:eqv1}, considering the fact that, in general, the proton detection was not selective on the relevant variables, but was mostly used to eliminate the background. 

\begin{table}[ht]
\begin{center}
\begin{ruledtabular}
\caption[]{Summary of experimental data on the ratio of positron to electron elastic scattering cross section off the nucleon.}
\begin{tabular}{|c|c|c|c|c|c|c|}
\hline
Ref.     & $\epsilon$ & $Q^2$     &  $R^{meas.}$  & $\frac{\Delta E^\prime}{E}$& $RC^{ex}$ & $RC^{th}$   \\
         &            & [GeV$^2$] &               &                      &            &              \\ \hline
\protect\cite{Mar68}& 0.972 & 0.689 & 0.992 $\pm$ 0.006 & 0.010 & -0.006 & -0.007\\ \hline
\protect\cite{Mar68}& 0.918 & 1.540 & 1.018 $\pm$ 0.016 & 0.010 & -0.015 & -0.018\\ \hline
\protect\cite{Mar68}& 0.831 & 2.440 & 1.068 $\pm$ 0.041 & 0.010 & -0.028 & -0.030\\ \hline
\protect\cite{Mar68}& 0.723 & 3.270 & 1.156 $\pm$ 0.122 & 0.010 & -0.045 & -0.044\\ \hline
\protect\cite{Mar68}& 0.999 & 0.204 & 1.011 $\pm$ 0.013 & 0.010 & -0.001 & -0.001\\ \hline
\protect\cite{Mar68}& 0.995 & 0.731 & 0.967 $\pm$ 0.043 & 0.010 & -0.002 & -0.003\\ \hline
\protect\cite{Mar68}& 0.953 & 3.790 & 1.038 $\pm$ 0.032 & 0.010 & -0.014 & -0.015\\ \hline
\protect\cite{Mar68}& 0.923 & 5.000 & 1.058 $\pm$ 0.057 & 0.010 & -0.020 & -0.021\\ \hline
\protect\cite{Yo62}& 0.874 & 0.010 & 0.998 $\pm$ 0.012 & 0.030 & -0.002 & -0.002\\ \hline
\protect\cite{Yo62}& 0.874 & 0.020 & 0.978 $\pm$ 0.016 & 0.030 & -0.002 & -0.003\\ \hline
\protect\cite{Yo62}& 0.874 & 0.020 & 1.008 $\pm$ 0.018 & 0.030 & -0.002 & -0.003\\ \hline
\protect\cite{Yo62}& 0.742 & 0.050 & 1.010 $\pm$ 0.032 & 0.030 & -0.006 & -0.007\\ \hline
\protect\cite{Yo62}& 0.094 & 0.190 & 1.066 $\pm$ 0.064 & 0.030 & -0.022 & -0.029\\ \hline
\protect\cite{Br65}& 0.737 & 0.140 & 1.010 $\pm$ 0.023 & 0.030 & -0.012 & -0.010\\ \hline
\protect\cite{Br65}& 0.291 & 0.760 & 1.114 $\pm$ 0.040 & 0.030 & -0.045 & -0.039\\ \hline
\protect\cite{Br65}& 0.291 & 0.760 & 1.207 $\pm$ 0.088 & 0.020 &    -   & -0.046\\ \hline
\protect\cite{Br65}& 0.376 & 0.620 & 0.981 $\pm$ 0.084 & 0.020 &    -   & -0.039\\ \hline
\protect\cite{Br65}& 0.466 & 0.600 & 1.038 $\pm$ 0.065 & 0.020 &    -   & -0.034\\ \hline
\protect\cite{Br65}& 0.237 & 0.480 & 1.091 $\pm$ 0.067 & 0.020 &    -   & -0.042\\ \hline
\protect\cite{Br65}& 0.681 & 0.270 & 1.010 $\pm$ 0.031 & 0.020 &    -   & -0.017\\ \hline
\protect\cite{An68}& 0.798 & 0.230 & 1.005 $\pm$ 0.024 & 0.100 & -0.014 & -0.006\\ \hline
\end{tabular}
\label{table:table1}
\end{ruledtabular}
\end{center}
\end{table}

\begin{table}[ht]
\begin{center}
\begin{ruledtabular}
\caption[]{Continue.}
\begin{tabular}{|c|c|c|c|c|c|c|}
\hline
Ref.     & $\epsilon$ & $Q^2$     &  $R^{meas.}$  & $\frac{\Delta E^\prime}{E}$ & $RC^{ex}$ & $RC^{th}$    \\
         &            & [GeV$^2$] &               &                      &            &               \\ \hline
\protect\cite{An68}& 0.553 & 0.465 & 1.006 $\pm$ 0.022 & 0.100 & -0.014 & -0.012\\ \hline
\protect\cite{An68}& 0.827 & 0.405 & 1.036 $\pm$ 0.025 & 0.100 & -0.022 & -0.006\\ \hline
\protect\cite{An68}& 0.709 & 0.620 & 0.990 $\pm$ 0.029 & 0.100 & -0.022 & -0.009\\ \hline
\protect\cite{An68}& 0.564 & 0.850 & 1.002 $\pm$ 0.045 & 0.100 & -0.022 & -0.014\\ \hline
\protect\cite{An66}& 0.697 & 0.640 & 1.008 $\pm$ 0.020 & 0.100 & -0.012 & -0.010\\ \hline
\protect\cite{Ca67}& 0.816 & 0.750 & 1.014 $\pm$ 0.039 & 0.020 & -0.039 & -0.018\\ \hline
\protect\cite{Ca67}& 0.730 & 1.000 & 1.025 $\pm$ 0.034 & 0.020 & -0.039 & -0.025\\ \hline
\protect\cite{Ba67}& 0.949 & 0.454 & 1.019 $\pm$ 0.032 & 0.010 & -0.007 & -0.009\\ \hline
\protect\cite{Ba67}& 0.784 & 1.366 & 0.981 $\pm$ 0.057 & 0.010 & -0.027 & -0.031\\ \hline
\protect\cite{Bo68}&   0   & 0.311 & 1.085 $\pm$ 0.018 & 0.010 & -0.049 & -0.058\\ \hline
\protect\cite{Bo68}&   0   & 1.246 & 1.180 $\pm$ 0.054 & 0.003 & -0.101 & -0.147\\ \hline
\protect\cite{Ha79}& 0.972 & 0.342 & 1.027 $\pm$ 0.060 & 0.060 &    -   & -0.002\\ \hline
\protect\cite{Ha79}& 0.972 & 0.450 & 1.041 $\pm$ 0.031 & 0.060 &    -   & -0.002\\ \hline
\protect\cite{Ha79}& 0.987 & 0.223 & 0.935 $\pm$ 0.040 & 0.045 &    -   & -0.001\\ \hline
\protect\cite{Ha75}& 0.972 & 0.343 & 0.974 $\pm$ 0.050 & 0.035 & -0.004 & -0.003\\ \hline
\protect\cite{Ha75}& 0.987 & 0.224 & 0.982 $\pm$ 0.050 & 0.035 & -0.002 & -0.002\\ \hline
\protect\cite{Ha75}& 0.972 & 0.449 & 1.044 $\pm$ 0.060 & 0.035 & -0.004 & -0.004\\ \hline
\protect\cite{Ca69}& 0.998 & 0.155 & 0.940 $\pm$ 0.049 & 0.130 &    -   &    -  \\ \hline
\protect\cite{Ca69}& 0.996 & 0.255 & 1.000 $\pm$ 0.052 & 0.130 &    -   &    -  \\ \hline
\protect\cite{Ca69}& 0.994 & 0.353 & 1.099 $\pm$ 0.077 & 0.130 &    -   &    -  \\ \hline
\protect\cite{Ca69}& 0.993 & 0.451 & 0.890 $\pm$ 0.121 & 0.130 &    -   &    -  \\ \hline
\end{tabular}
\label{table:table1b}
\end{ruledtabular}
\end{center}
\end{table}

\begin{table}[ht]
\begin{center}
\begin{ruledtabular}
\caption[]{Continue.}
\begin{tabular}{|c|c|c|c|c|c|c|}
\hline
Ref.     & $\epsilon$ & $Q^2$     &  $R^{meas.}$  & $\frac{\Delta E^\prime}{E}$ &  $RC^{ex}$ & $RC^{th}$     \\
         &            & [GeV$^2$] &               &                      &            &               \\ \hline
\protect\cite{Ca69}& 0.991 & 0.551 & 0.753 $\pm$ 0.140 & 0.130 &    -   &    -  \\ \hline
\protect\cite{Ca69}& 0.989 & 0.652 & 0.503 $\pm$ 0.251 & 0.130 &    -   &    -  \\ \hline
\protect\cite{Ca69}& 0.986 & 0.753 & 1.053 $\pm$ 0.436 & 0.130 &    -   &    -  \\ \hline
\protect\cite{Ca69}& 0.984 & 0.849 & 0.927 $\pm$ 0.486 & 0.130 &    -   &    -  \\ \hline
\protect\cite{Ca69}& 0.999 & 0.145 & 1.004 $\pm$ 0.038 & 0.130 &    -   &    -  \\ \hline
\protect\cite{Ca69}& 0.999 & 0.245 & 0.947 $\pm$ 0.044 & 0.130 &    -   &    -  \\ \hline
\protect\cite{Ca69}& 0.998 & 0.346 & 0.963 $\pm$ 0.087 & 0.130 &    -   &    -  \\ \hline
\protect\cite{Ca69}& 0.998 & 0.444 & 1.151 $\pm$ 0.197 & 0.130 &    -   &    -  \\ \hline
\protect\cite{Ca69}& 0.997 & 0.543 & 0.775 $\pm$ 0.187 & 0.130 &    -   &    -  \\ \hline
\protect\cite{Ca69}& 0.997 & 0.645 & 1.101 $\pm$ 0.417 & 0.130 &    -   &    -  \\ \hline
\protect\cite{Ca69}& 0.996 & 0.745 & 0.903 $\pm$ 0.483 & 0.130 &    -   &    -  \\ \hline
\end{tabular}
\label{table:table1c}
\end{ruledtabular}
\end{center}
\end{table}

Let us review the main features of these data:
\begin{enumerate}

\item Ref.~\cite{Ca69} reports on the experiment with $\mu^+/\mu^-$ beams at AGS (Brookhaven), where not only charge asymmetry, but also deviation from linearities of the Rosenbluth plot were measured in the range $0.15\le Q^2\le 0.85$ GeV$^2$. In this case the radiative corrections are suppressed by the mass of the lepton, they were considered independent of the lepton charge, and were note applied.

\item In Refs.~\cite{An66,An68} the ratio $R$ was measured at Cornell synchrotron with 0.8 and 1.2 GeV $e^+/e^-$ beam energy, spread over 10\% magnetic collimator slit. The beam energy was reconstructed from the lepton scattered angle and proton recoil angle measured in coincidence. The elastic events were selected using the reconstructed beam energy $E'$, with the cut $\Delta E^\prime_{rec}/E<0.15$. $Q^2$-dependent radiative corrections were applied following Ref.~\cite{Ye61}. It has to be stressed that the numbers given in Table III of Ref.~\cite{An68} for $E=1.2$ GeV are sometimes inconsistent among each other and with Fig. 10. Therefore we used Fig. 10 to obtain correct values of the ratio at $Q^2=0.27-0.54$ GeV$^2$ and $Q^2=0.70-1$ GeV$^2$.

\item  Ref.~\cite{Ca67} reported on a similar measurement at Cornell synchrotron using 1.7 GeV $e^+/e^-$ beam. In this case, however, the (Gaussian) beam energy spread, $\sigma_{Beam}$, was considerably smaller, $\sigma_{Beam}=$42 MeV. This suggests $\Delta E^\prime_{rec}/E<0.075$, considering 3$\sigma_{Beam}$ cut on the reconstructed beam energy, as in Refs.~\cite{An66,An68}. The radiative corrections, as large as 4\%, were calculated from Ref.~\cite{Me63}, although this number would imply the proton momentum resolution of $5\times10^{-4}$ and $4\times10^{-3}$ for $Q^2=0.75$ and $Q^2=1$ GeV$^2$, respectively. These values are in strong contrast with the beam energy spread, which we will use for our radiative correction estimate.

\item $e^+/e^-$ elastic scattering on protons was measured at DESY as reported in Ref.~\cite{Ha75} (open stars, gray). Radiative corrections according to Ref.~\cite{Me63} were applied to the data. The consistency with Ref. \cite{Mo69} was checked. As the parameter $\Delta E^\prime/E$ was not given explicitly in this paper, we took the value from Ref.~\cite{Gal72} where the inelasticity cut used was chosen to be  $W<1.05$ GeV ($W$ is the invariant mass). The resulting corrections are negligible with respect to statistical uncertainties. This measurement was repeated in Ref.~\cite{Ha79} on $^{12}C$ and $^{27}Al$ targets.

\item In Ref.~\cite{Br65} $e^+/e^-$ elastic scattering off the proton was measured at SLAC in two different setups. The first experiment detected scattered lepton and proton in coincidence as in Refs.~\cite{An66,An68,Ca67}. However, the point at highest $Q^2$ was strongly contaminated by meson electro- and photoproduction products (almost half of measured events). To remeasure this point in a clean manner the second experiment was performed. We drop this contaminated point and keep only the second experiment data for this kinematics.
$\Delta E^\prime/E$ values were deduced from $\Delta E_4$ given in Ref.~\cite{Atkinson}.
The second experiment used a single arm magnetic spectrometer to detect the scattered lepton. The applied radiative corrections were tabulated. The value $\Delta E^\prime/E=0.018$ is coherent with the measured spectrum as well as with the values of the correction.

\item In Ref.~\cite{Ba67} $e^+/e^-$ elastic scattering off the proton was measured at DESY using 2.24 GeV beam energy. The beam energy spread was 0.6\%, resulting in 1.5\% elastic peak width. Events with $\Delta E^\prime/E<0.03$ had been selected.
Radiative corrections according to Ref.~\cite{Me63} were applied to the data.

\item In Ref.~\cite{Bo68} $e^+/e^-$ elastic scattering off the proton was measured at $\theta=180^\circ$,
where, according to Gourdin \cite{Go66}, a larger effect of TPE was expected. Here the backward scattered electron was detected in coincidence with the forward proton. The main author updated the data in his PhD thesis, published three years later. In this thesis a non-negligible background as a function of the inelasticity cut was pointed out for the high $Q^2$ point. Therefore we selected the latest result for most stringent inelasticity cut. The inelasticity cut was applied to the beam energy spectrum, reconstructed from the angles of the outgoing particles. The coplanarity spectrum showed very little background.
Custom radiative corrections, from Ref.~\cite{Me63}, in their C-odd part, had been applied to the data, taking into account the effects of solid angles. After comparing the calculations \cite{Me63} and \cite{Ku06} for the corresponding cut $\Delta E^\prime/E^\prime=0.10$, we renormalized our value to the published value, making the assumption that the geometry has the same effect in the two calculations.

\end{enumerate}

\subsection{Inelastic scattering}

The signature of TPE was searched also in the inelastic lepton-nucleon scattering. In this case, the final state hadronic system remains undetected and can be only characterized by its invariant mass $W$. 

In case on a nuclear target, $A(Z,N)$, in the quasi-elastic region this reaction can be approximately described as the incoherent sum of elastic scattering off individual nucleons, which is reasonable at sufficiently large $Q^2$ values:
\be
\sigma^{exp}_{e^{\pm}A}=Z\sigma^{exp}_{e^{\pm}p}+N\sigma^{exp}_{e^{\pm}n}
\label{eq:eqA}
\ee
Since the real photon emission from the neutron
is unlikely, the neutrons do not contribute to the asymmetry, more exactly to the part which is due to interference between electron and target emission. On the other hand, neutrons do contribute to the hard box, as they have non zero FFs (although such contribution is expected to be smaller than for protons). We take this into account by averaging the asymmetry in case of nuclear target:
$A^{odd}_{A}=Z A^{odd}/(Z+N)$, where $A^{odd}$ is the free proton asymmetry.

\begin{figure}
\begin{center}
\includegraphics[bb=1cm 6cm 20cm 23cm, scale=0.4]{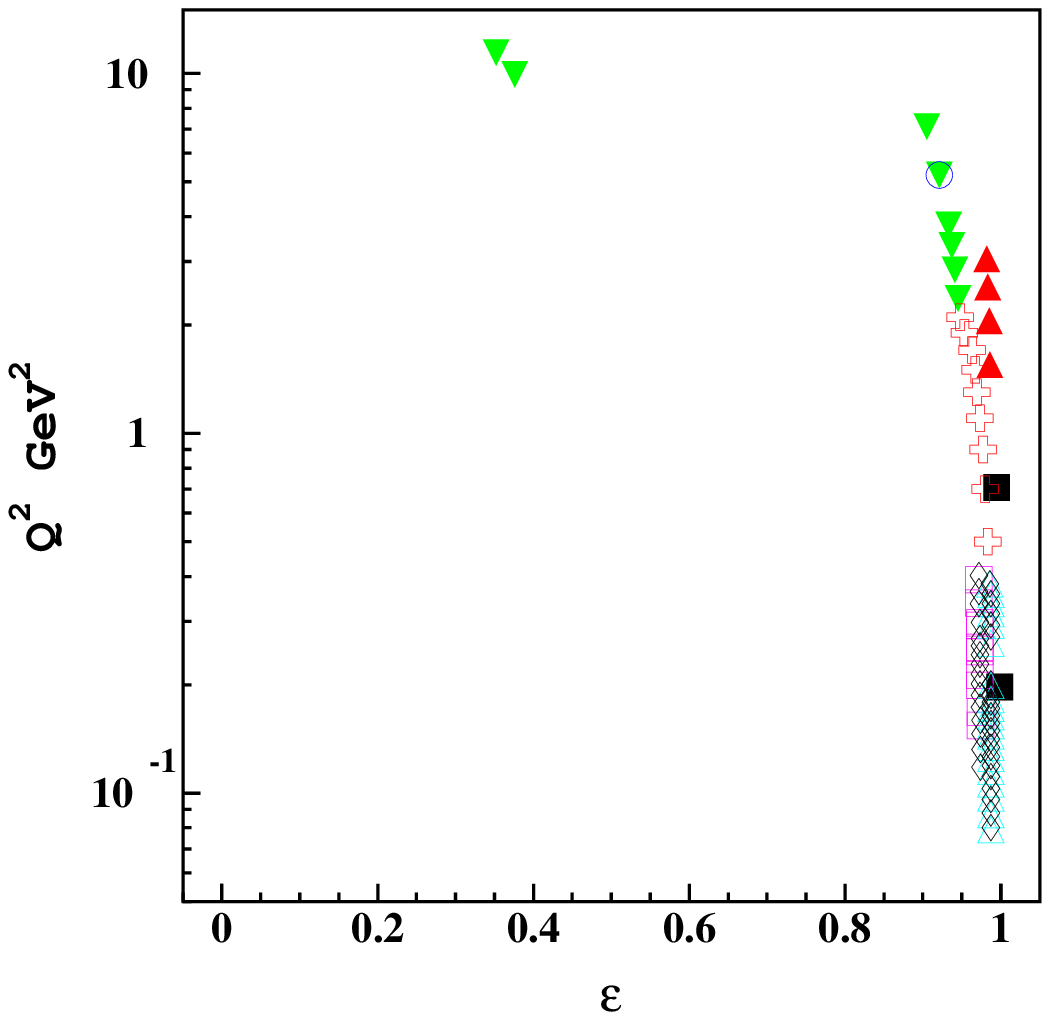}~
\includegraphics[bb=1cm 6cm 20cm 23cm, scale=0.4]{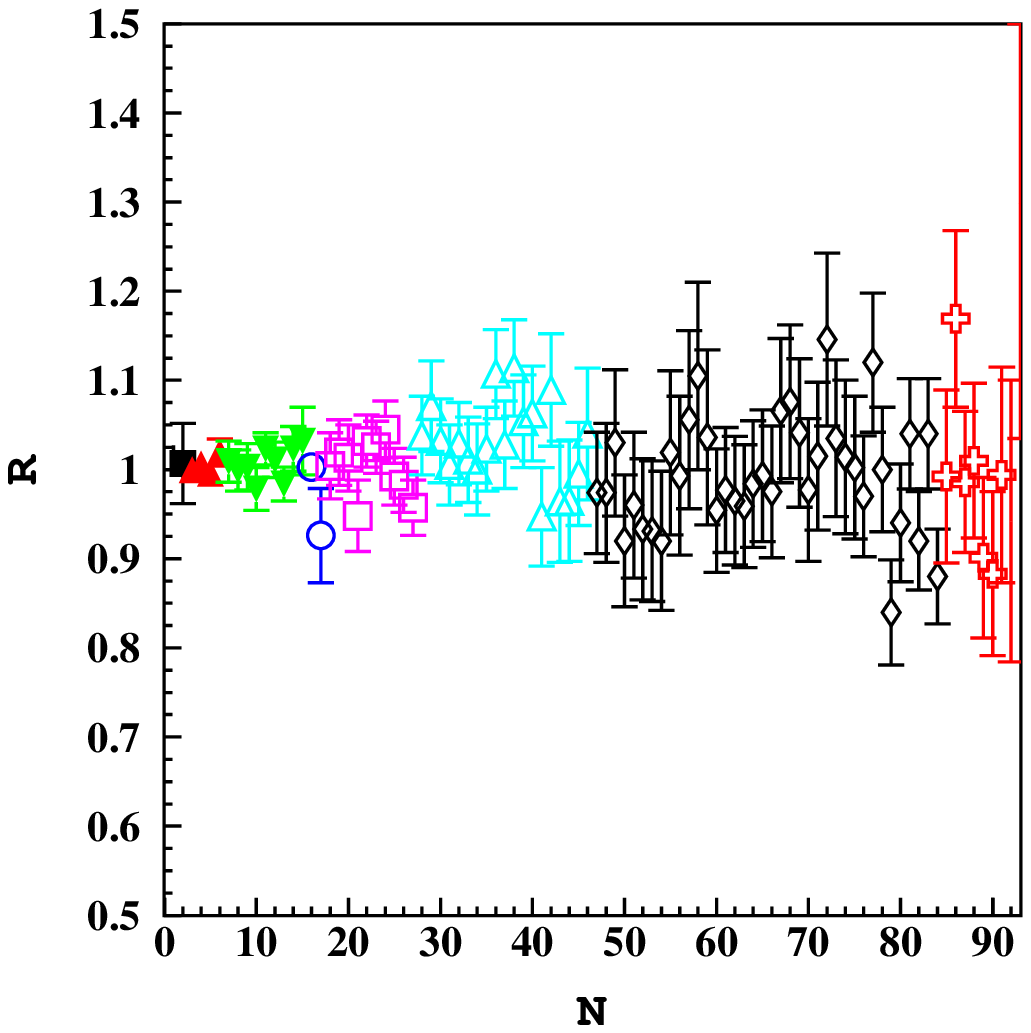}
\caption{\label{fig:r_in_data} (Color online) Kinematics and absolute values of the published inelastic cross section ratios,  $R=\sigma(e^+)/\sigma(e^-)$. Here $N$ is a number attributed to each point, following a sequential order from Refs. 
\protect\cite{Mar68} (solid squares, black),
\protect\cite{Fa76} (triangles, red),
\protect\cite{Ro76}-H target (triangles down, green),
\protect\cite{Ro76}-D target  (open circles, blue),
\protect\cite{Ha79}-$^{12}$C target (open squares, magenta),
\protect\cite{Ha79}-$^{27}$Al target (open triangles, cyan),
\protect\cite{Ha76} (open lozenges, black),
\protect\cite{Jo74} (open crosses, red).}
\end{center}
\end{figure}

They are summarised in Fig.~\ref{fig:r_in_data}, where the $Q^2$ and $\epsilon$ values covered by the data are shown (left) and the ratio of inelastic $e+/e^-$ cross sections is plotted as a function of a sequential number attributed to the points (right). No radiative corrections were applied to the data.

Let us do a brief summary of these experiments, referring to Fig. \ref{fig:r_in_data}. 
\begin{itemize}

\item In Ref. \cite{Fa76} (triangles, red) the main result, $R=1.0027\pm 0.0035$ was obtained as an average of four measurement in the range $1.2<Q^2<3.3$ GeV$^2$ and $2<\nu<9.5$, after insuring that there was no systematic trend of the data in the spanned kinematical range. This  measurement is quite precise, therefore especially interesting for our discussion. The difference for electron and positron cross sections was very small. The lepton scattering angle was $\theta=8^{\circ}$, and the measurements correspond to large $\epsilon\sim 0.98$.
\item Ref. \cite{Ro76} (triangles down, green) reports on measurements on hydrogen  (triangles down, green)and deuterium (open circles, blue) up to $Q^2$=15 GeV$^2$. The ratio is consistent with unity, within errors of a few percent. Specific settings of the spectrometer allowed to measure different charges, alternatively. 
\item Inelastic scattering on protons at DESY has also been reported in Ref. \cite{Ha76} (open lozenges, black) for $\theta$ =9 and $13^{\circ}$, giving a value of the ratio compatible with one, within an error of 4 and 5$\%$. The inelastic region from $1.2<W<3.4$ was covered by several measurements in which no systematic trend was observed. 
In the work in Ref. \cite{Ha79} $e^{\pm}$ inelastic scattering on $^{12}C$ (open squares, magenta),and $^{27}Al$ (open triangles, cyan) was investigated. The final result $R=1.005\pm 0.027$, was obtained in the region of momentum transfer $0.08<Q^2<0.45$ GeV$^2$ and invariant mass $0.95\le W\le 3.3$ GeV of the hadronic system. The final result has been averaged from several measurements, after verifying that no dependence on the momentum transfer, on the inelasticity and on the charge of the target appeared in the limit of the experimental error. 
\item A measurement on deep inelastic scattering \cite{Jo74} (open crosses, red), done at AGS (Brookhaven), in the range for $Q^2<2.1$ GeV$^2$ and $\nu<$ 5 GeV, concluded that TPE amplitudes contribute less that 0.17$\%$. 

\end{itemize}

In the kinematic conditions of the experiment \cite{Ha79}, one should note that the scattering angle is very small ($\epsilon\sim 1$), as well as $Q^2$. In these kinematical conditions soft photon emission is very small, inducing very small asymmetry. On the other hand, one expects multiphoton exchange effects by the strong Coulomb field of the nuclei. To give an order of magnitude, for the targets considered here which are relatively light, Ref. \cite{Ku09} predicts an effect of $\sim 2\%$ at E=3 GeV and for scattering angle $\theta=9^0$ from multiphoton exchange calculated in elastic kinematics.

In the inelastic case the corrections from Ref. \cite{Ku08} are not applicable and therefore the data were compared to the lowest order solution, $R=1$. The 93 points from the inelastic data are consistent with $R=1$ and $\chi^2=0.8$.

\section{Discussion}

\subsection{Comparison with theory}

We report in  Tables \ref{table:table1}, \ref{table:table1b}, and \ref{table:table1c} the results of the different experiments for elastic electron and positron scattering off the nucleon, together with the values of the relevant kinematical variables $Q^2$, and $\epsilon$ as well as the radiative corrections applied to the data and calculated from \cite{Ku06}.

In order to verify the effect of standard radiative corrections, and to compare to the theoretical predictions, one has to subtract the applied radiative corrections from the data, taking into account that different ansatz for radiative corrections were used in different experiments.

A quantitative comparison of the data with theoretical calculation of Ref. \cite{Ku08}
has been done. In order to unfold the role of each variable, let us define the $\chi^2$:
as
\be
\chi^2(Q^2,\epsilon,\Delta E) =\left [\frac{R^{raw}_i-R^{th}(Q^2,\epsilon,\Delta E)}{\Delta R^{raw}(Q^2,\epsilon,\Delta E)}\right ]^2,
\label{eq:eqth}
\ee
where $R^{raw}\pm \Delta R^{raw}$ are the experimental data including all corrections besides radiative corrections, and $R^{th}$ is built from Eqs. (\ref{eq:ratio}) and (\ref{eq:eqv1}), for the corresponding experimental conditions. We do not attribute any error to the theoretical value.
Combining the 53 elastic data points  we find  $\chi^2=1$. Although, it does not result from a minimization procedure, as no free parameters have been used, this constitutes a check of the validity of the theoretical hypothesis. 

\begin{figure}
\begin{center}
\includegraphics[scale=0.4]{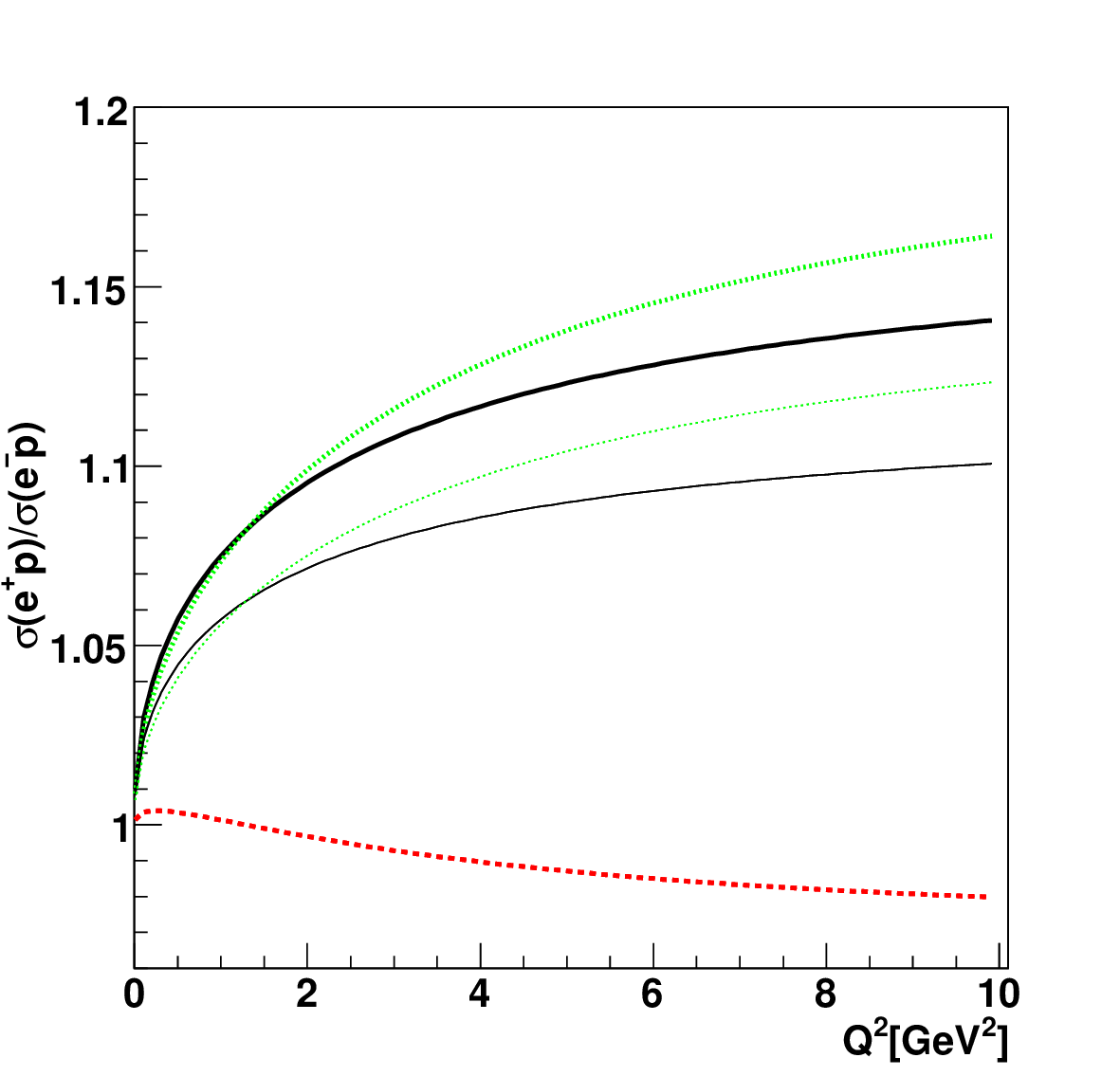}~%
\includegraphics[scale=0.4]{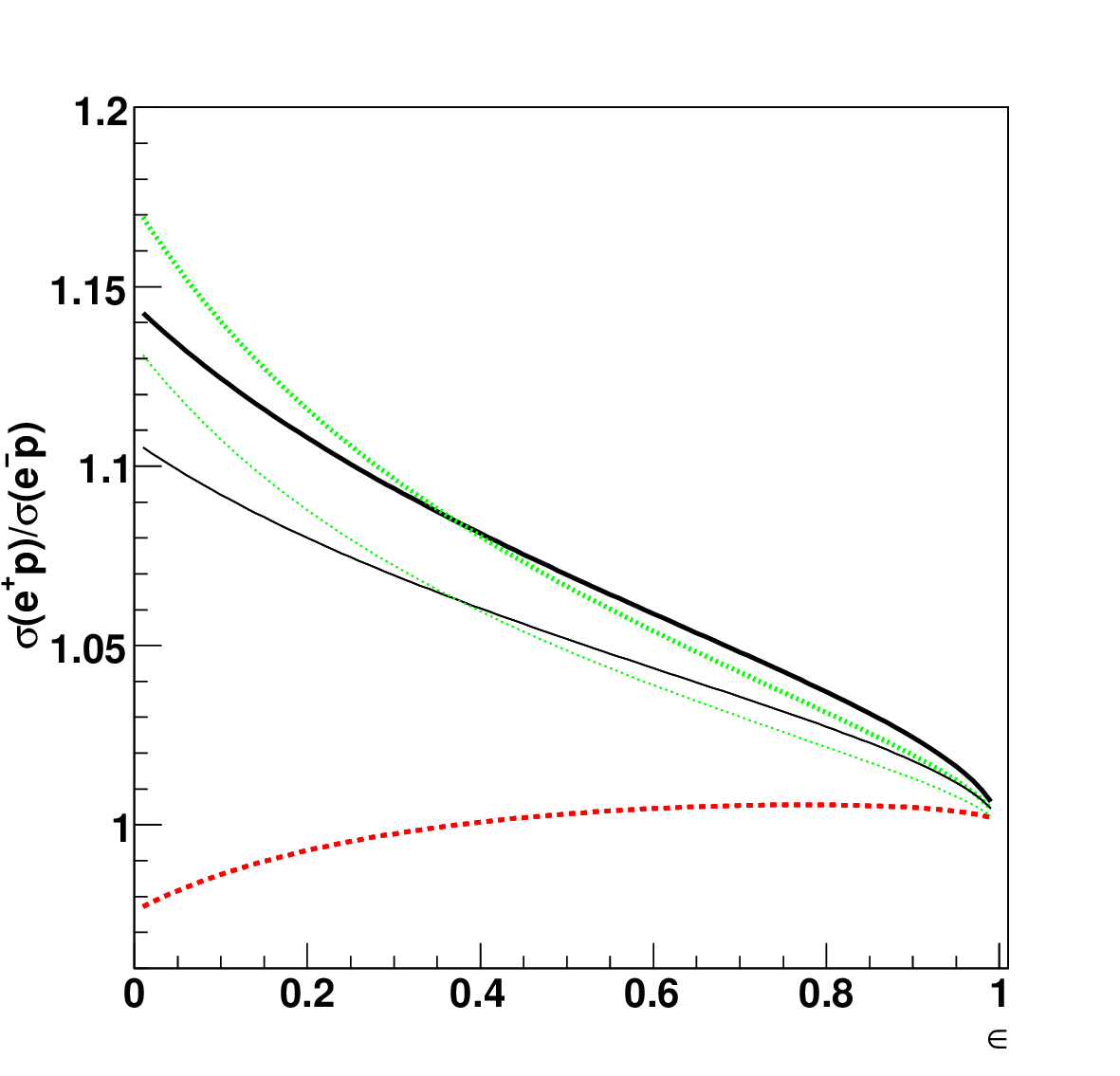}
\caption{(Color online) Ratio of cross sections $R=\sigma(e^+p)/\sigma(e^-p)$, as a function of $Q^2$ (left) ($\epsilon=0.2$) and $\epsilon$, ($Q^2$=3 GeV$^2$), $\Delta E^\prime/E=0.03$ (thick lines) and $\Delta E^\prime/E=0.01$ (thin lines) from Ref. \protect\cite{Ku08}: total contribution (solid line, black), soft contribution (dotted line, green). The hard contribution (dashed line, red) does not depend on the cut.}
\label{fig:figcut}
\end{center}
\end{figure}

After the theory has been validated by the comparison to the experiment we can explore the main kinematic dependences.
The behaviour of the ratio $R$ as a function of $Q^2$ and $\epsilon$ is shown in Fig.~\ref{fig:figcut} for two values of the inelasticity cut: $\Delta E^\prime/E=0.03$ (thin lines) and $\Delta E^\prime/E=0.01$ (thick lines).
The {\it hard box} does not depend on the inelasticity cut (red, dashed line). The solid lines correspond to the full contribution, and the dotted lines to the soft contribution. The deviation from $R=1$ is increasing at large $Q^2$ and small $\epsilon$. The soft contribution is increasing with $\Delta E^\prime \to 0$. The hard box contribution, although less sizable, has similar trends, but has opposite sign reducing the overall ratio.

Therefore, we can conclude that the deviation from unity of the measured ratios have to be attributed to radiative corrections, mostly due to soft photon emission.

\subsection{Comparison to $G_E/G_M$ data}

As mentioned in the Introduction, FFs obtained by polarization transfer and Rosenbluth techniques deviate more and more as $Q^2$ becomes large.  Radiative corrections are large in unpolarized measurements, whereas, in first approximation, they cancel in the porarised cross sections ratio. Typically the Rosenbluth type experiments have been corrected with first order radiative corrections from \cite{Me63,Mo69,MT00}, whereas no radiative correction to the polarization-type experiment has been applied.

It has been argued that TPE may reconcile these measurements~\cite{Gu03,Bl05,Af05}.
To evaluate the validity of this statement, let us assume that the difference between the reduced elastic cross sections deduced from polarized ($\sigma^P$) and unpolarized ($\sigma^R$) measurements is fully due to the hard box contribution. This can be expressed as follows:
\be
C_{2\gamma}=\sigma^R-\sigma^P=
(G_M^{P})^2\Biggl[\tau+\frac{\epsilon}{\mu^2} \Bigl(\frac{\mu G_E^{P}}{G_M^{P}}\Bigr)^2\Biggr] -
(G_M^R)^2 \Biggl[\tau + \frac{\epsilon}{\mu^2} \Bigl(\frac{\mu G_E^R}{G_M^R}\Bigr)^2\Biggr]
\label{eq:tpe_el}
\ee
where $\mu$ is the proton magnetic moment, $G_{E,M}^{P,R}$ are electric (E), magnetic (M) FFs of the proton obtained by means of polarization transfer (P) and Rosenbluth (R) techniques and $C_{2\gamma}$ represents the additional  contribution to the cross section due to an unknown TPE contribution.

First  we use the fact that the experimental data on the ratio of FFs can be simply parametrized  for Rosenbluth measurements as $\mu G_E^R/G_M^R=1$. Then we assume that TPE effect vanishes at $\epsilon=1$, as suggested by~\cite{Chen00}. Then, from Eq.~\ref{eq:tpe_el} we obtain:
\be
C_{2\gamma} = (\epsilon-1)\tau(G_M^R)^2\frac{1-(\mu G_E^{P}/G_M^{P})^2}{\mu^2\tau+(\mu G_E^{P}/G_M^{P})^2}
\label{eq:c2g}
\ee
The absence of non-linearities of the Rosenbluth plots~\cite{ETG05,Tv06} is consistent with $(\epsilon-1)$ form of the TPE $\epsilon$-dependence (\ref{eq:c2g}).

\begin{figure}
\begin{center}
\includegraphics[bb=1cm 6cm 20cm 23cm, scale=0.4]{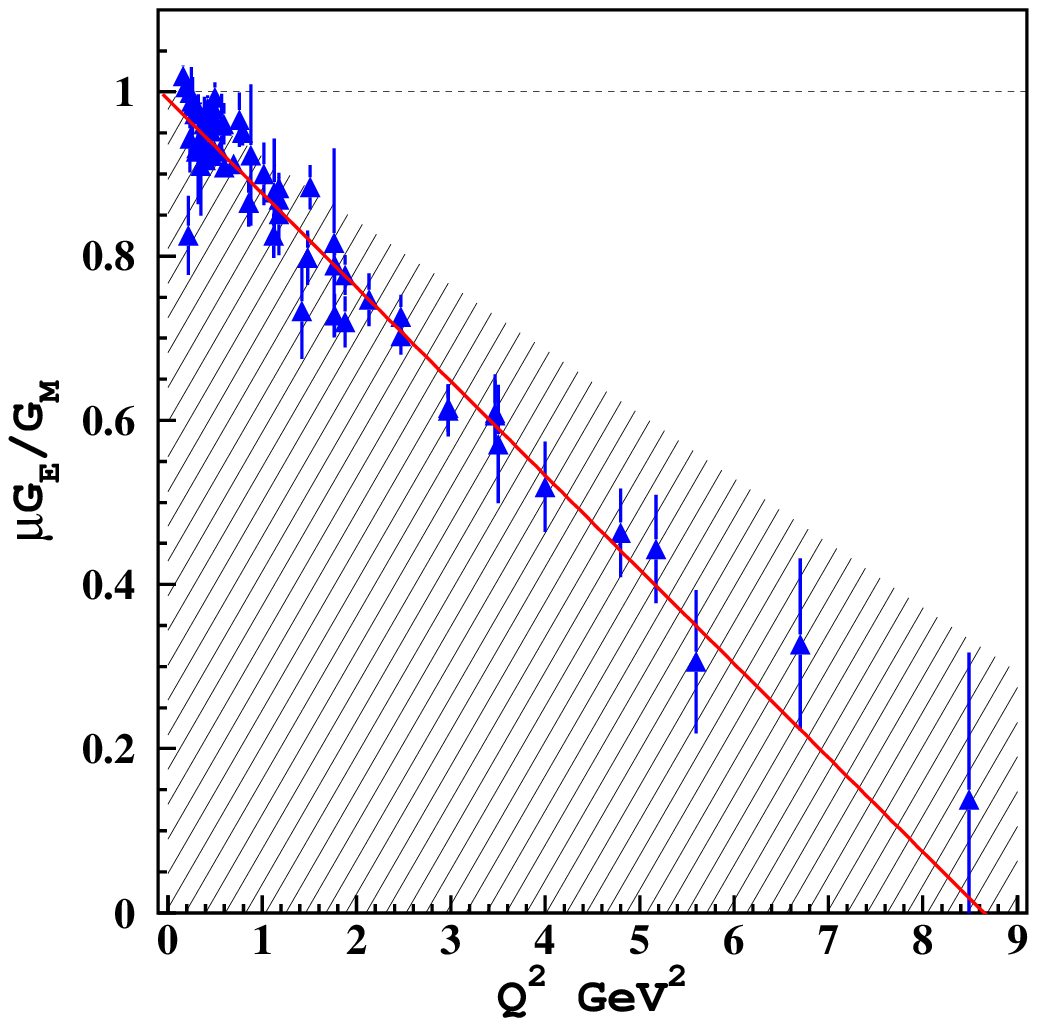}
\caption{(Color online) Ratio of proton FFs $\mu G_E^{PT}/G_M^{PT}$, as a function of $Q^2$. The line is the linear fit to the data and the hatched area is excluded by lepton-charge asymmetry measurement data at significance level of 0.05.}
\label{Fig:gegm}
\end{center}
\end{figure}

We can, therefore relate $C_{2\gamma}$ from Eq. (\ref{eq:c2g}) with the correction to the lepton-charge asymmetry by:
\be
A^{odd}_{2\gamma}=\frac{C_{2\gamma}}{\sigma^R -C_{2\gamma}}=
(\epsilon-1)\Biggl[
1-\epsilon+\Bigl(1+\frac{\epsilon}{\mu^2\tau}\Bigr)\frac{\mu^2\tau+(\mu G_E^{P}/G_M^{P})^2}{1-(\mu G_E^{P}/G_M^{P})^2}
\Biggr]^{-1}
\label{eq:aodd_P}
\ee
Using the expression from Eq.~\ref{eq:aodd_P} we can test the hypothesis that the difference between polarization transfer and Rosenbluth extraction of $\mu G_E/G_M$ ratio is fully due to TPE contribution. Assuming significance level of 0.05 we found the $Q^2$-slope of $\mu G_E/G_M$ ratio that would invalidate the hypothesis. The region excluded by this procedure is shown in Fig.~\ref{Fig:gegm} by hatched area. Within our assumptions at least the data for $Q^2>2$ GeV$^2$ cannot be described by TPE contribution.

\section{Conclusions}

We have reanalyzed the existing data on lepton-charge asymmetry in elastic and inelastic scattering off the nucleon. We have compared the data to a model calculation \cite{Ku08}, which does not contain free parameters, and a good compatibility was found on the basis of a point-to-point quantitative analysis. Within this analysis, the C-odd soft contribution, arising from the interference between electron and target bremsstrahlung, gives the main contribution to the observed asymmetry in the elastic case.
    
Note that, if we apply to the data radiative corrections from a different prescription, 
Ref. \cite{MT00}, the values of the corrections coincide at the level of $1.5\%$, and the effect on the corrected ratio is in average of the order of a percent, not affecting our conclusions. This is expected from the fact that both Ref. \cite{MT00} and Ref. \cite{Ku08} are first order calculations, the first one taking into account only infrared divergent part of TPE, whereas the second one predicts a small hard box TPE contribution.

The ratio issued from inelastic scattering data is consistent with unity. 
 
A good understanding of the validity of OPE is very important for different applications. Given that our analysis shows the limited precision of the existing data, more precise measurements from CLAS~\cite{clas_tpe}, OLYMPUS~\cite{olympus_tpe} and VEPP3\cite{Gramolin:2011tr} collaborations are expected to bring stronger constrains.

\section{Acknowledgments}
We acknowledge V. V. Bytev for essential help and for useful discussions. Thanks are due to G.I.Gakh for careful reading and interesting remarks. This work was supported in part by the GDR PH-QCD. E.A.K. and Yu. M.B. acknowledge the grant INTAS N 05-1000008-8528 for financial support.

\end{document}